\documentclass[rnote]{aa}

\usepackage{graphicx}
\usepackage{amsmath}
\usepackage[dvipsnames]{xcolor}
\usepackage{multirow}
\usepackage{txfonts}
\usepackage{natbib}
\bibpunct{(}{)}{;}{a}{}{,} 
\usepackage{booktabs} 
\usepackage{url}
\usepackage{hyperref}
\hypersetup{colorlinks = true,
            linkcolor = blue,
            urlcolor  = blue,
            citecolor = blue,
            anchorcolor = blue
            }
\begin{document} 
    \titlerunning{Mocking IFU observations from simulations}
    \authorrunning{Cornejo-Cárdenas et al.}
   
    \title{Mocking IFU observations and metallicity diagnostics from cosmological simulations}
    
    \author{Anell Cornejo-Cárdenas\inst{1,2,3,}\thanks{E-mail : \url{avcornejo@uc.cl}}, 
    Emanuel Sillero\inst{1,3}, 
    Patricia B. Tissera\inst{1,3}, 
    Médéric Boquien\inst{4}, 
    José Vilchez\inst{5},
    Gustavo Bruzual\inst{6},
    Paula Jofré\inst{3,7}
    }
      
    \institute{Institute of Astronomy, Pontificia Universidad Católica de Chile, Av. Vicuña Mackenna 4860, Santiago, 7820436, Chile. 
    \and
    Centro de Astro-Ingeniería, Pontificia Universidad Católica de Chile, Santiago, Chile
    \and
    Nucleus Millenium ERIS, Santiago, Chile.
    \and 
    Université Côte d'Azur, Observatoire de la Côte d'Azur, CNRS, Laboratoire Lagrange, F-06000 Nice, France.
    \and
    Instituto de Astrofísica de Andalucia, Glorieta de la Astronomía s/n, 18008 Granada, Spain.
    \and 
    Instituto de Radioastronomía y Astrofísica (IRYA), UNAM, Campus Morelia, Michoacán, C.P. 58089, México.
    \and
    Instituto de Estudios Astrofísicos, Facultad de Ingeniería y Ciencias, Universidad Diego Portales, Santiago de Chile.
    }

   \date{Received 12 September 2024; Accepted 27 May 2025}

    \abstract
    {Hydrodynamic simulations are powerful tools for studying galaxy formation. However, it is crucial to test and improve the sub-grid physics underlying these simulations by comparing their predictions with  observations. 
    To this aim, observable quantities can be derived for simulated galaxies, enabling the analysis of simulated properties through an observational approach.}
    {Our goal is to develop a new numerical tool capable of generating synthetic emission line spectra from spatially resolved regions in simulated galaxies, mocking Integral Field Unit (IFU) observations.}
    {Synthetic spectra of simulated galaxies are produced by integrating the software CIGALE with the outcomes of hydrodynamical simulations. We consider contributions of both stellar populations and nebular emission. The nebular emission lines in the spectra are modeled by considering only the contributions from the simulated star-forming regions. Our model considers the properties of the surrounding interstellar medium to estimate the ionizing parameters, the metallicity, the velocity dispersion and the electron density.}
    {We present the new numerical tool PRISMA.
    Leveraging synthetic spectra generated by our model, PRISMA successfully computed and recovered the intrinsic values of the star formation rate and gas-phase metallicity in local regions of simulated galaxies. Additionally, we examine the behavior of metallicity tracers such as N2, R23, O3N2, N2O2, recovered by PRISMA and propose new calibrations based on our simulated result. These findings show the robustness of our tool in recovering the intrinsic properties of simulated galaxies through their synthetic spectra, thereby becoming a powerful tool to confront simulations and observational data.}
    {}


    \maketitle

\section{Introduction}

In recent decades, Integral Field Unit (IFU) spectrographs, characterized by the ability to capture spectra from spatially resolved regions, have yielded invaluable datasets that significantly contribute to a deeper comprehension of the processes governing the formation and evolution of galaxies. Prominent examples of such IFU surveys include the Calar Alto Legacy Integral Field Area study \citep[CALIFA;][]{Sanchez2012}, the Sydney-AAO Multi-object IFS survey \citep[SAMI;][]{Croom2012}, the Mapping Nearby Galaxies at Apache Point Observatory \citep[MaNGA;][]{Bundy2015} and the Physics at High Angular Resolution in Nearby Galaxies \citep[PHANGS-MUSE;][]{Emsellem2022}. The growth in the data quality and scope of IFU surveys shows no sign of slowing, with new instruments staged to begin observing in the near future such as the Local Volume Mapper \citep[LVM;][]{Drory2024} and the Generalising Edge-on galaxies and their Chemical bimodalities, Kinematics, and Outflows out to Solar environments survey \citep[Geckos;][]{Van_de_Sande2024}. 

Leveraging IFU emission line spectra, astronomers can spatially resolve the physical properties of individual galaxies, as exemplified in studies by \citet{Barrera-Ballesteros2016, Sanchez2018b, Baker2023}, among others. In particular, these authors resorted to IFU data to investigate the local star formation rates (SFR) and chemical abundances, or metallicities\footnote{The metallicity (Z) is defined as the ratio between the mass of  elements heavier than hydrogen and helium in a given baryonic component and its mass.} (Z), across different galaxies with a variety of stellar mass and morphology.
 
With the advent of modern technology and the increasing number of statistical studies, scaling relations have been confirmed between different properties of galaxies, such as their mass, size, luminosity and colors \citep[e.g.,][]{Brinchmann2004, Tremonti2004, Daddi2007, Kewley_Ellison2008, Peng2010, Maier2014, Zahid2014, Renzini2015}. One example of these scaling relations is the mass-metallicity relation (MZR), which connects the gas-phase metallicity with stellar mass of the galaxies. Typically, these relations are estimated using integrated galaxy quantities, reflecting their overall properties. However, they may also be shaped by processes taking place on sub-galactic scales \citep{Boardman2022, Baker2023}. Consequently, delving into the properties of local regions within galaxies offers a pathway to better understanding the origin of the integrated relations.

Several studies based on IFU observations have been used to analyze the MZR at kiloparsec (kpc) scales \citep[e.g.,][]{Barrera-Ballesteros2016, Sanchez2021a, Yao_Yao2022}, confirming the existence of the local analogous relation known as the resolved MZR (rMZR). Numerical simulations have also been employed to study this relationship. \citep{TrayfordSchaye2019} investigated some scaling relations, including the rMZR, by analyzing the intrinsic properties of simulated galaxies selected from the EAGLE project \citep{Schaye2015}. Their study shows a very good agreement with observations from MaNGA and CALIFA surveys across various redshifts. However, in this work the intrinsic properties of the simulated galaxies are directly compared with the corresponding observed ones. An improvement over this could be achieved by combining simulations with spectral synthesis models to derive  synthetic spectra of simulated galaxies \citep[e.g.,][]{Tissera1997}. 

Forward modeling techniques can generate synthetic observations of simulations, facilitating a fair comparison between observed and simulated data and enabling consistent analysis with observational methods. This methodology has been previously employed to generate virtual or synthetic observations of simulations
\citep[e.g.,][]{Torrey2015, Snyder2015, Trayford2015, Bottrell2017, Trayford2017, Rodriguez-Gomez2018, Schulz2020, Nanni2022a, Garg2023, Hirschmann2023}. 
In particular, \citet{Nanni2022a} produced synthetic data cubes of IllustrisTNG galaxies \citep{Nelson2019} using real calibrated MaNGA stellar spectra \citep[MaStar stellar library;][]{Yan2019, Maraston2020, Abdurrouf2022}. By assembling synthetic spectra based on spectra obtained with fibers from the MaNGA spectrograph, they were able to incorporate noise into the synthetic spectra, replicating the signal-to-noise characteristics of real MaNGA galaxy observations. This numerical tool represents a major advancement in synthetic IFU data, proving a reasonable match to real data.  

Emission lines in galaxy spectra serve as tracers of their star formation activity and chemical abundances \citep{Kennicutt1983, Calzetti2013}. This makes them a valuable tool for estimating local properties such as metallicities, SFR and star formation histories (SFH). Specifically, the methods for estimating gas-phase metallicities based on  emission lines are: (i) the direct method (which is based on electron temperatures), (ii) use of recombination lines, (iii) empirical calibrations, and (iv) theoretical calibrations. The metallicity estimate is strongly dependent on the method used to calculate it and the calibrations involved \citep{Kewley_Ellison2008}. Hence, this is an extra difficulty when using observed chemical abundances to confront galaxy formation models. A plausible route is to employ forward modeling techniques to generate synthetic spectra with emission lines from simulated galaxies and make a three-fold comparison between intrinsic (from simulated models), predicted (from synthetic spectra) and observed abundances from IFU surveys. 

The main goal of this paper is to generate synthetic spectra with emission lines for comparing synthetic gas-phase abundances with observed counterparts. In this work, we introduce a new numerical tool, called Producing Resolved and Integrated Spectra from siMulated gAlaxies (PRISMA), designed to produce synthetic spectra by incorporating both stellar contributions and nebular emissions. A fundamental aspect of this tool, distinguishing it from previous works \citep[e.g.,][]{Trayford2017, Nanni2022a, Garg2023, Hirschmann2023} is that nebular emission is estimated from the simulated properties of the gas surrounding recently formed stellar populations. This aspect allows the modeling and analysis of parameters that characterizes the ionized interstellar medium (ISM), such as the ionization parameter (U), utilizing intrinsic properties directly provided by the simulation. In this paper, we describe  the main characteristics of PRISMA, the procedure developed to generated IFU-like datacubes of a simulated sample and apply PRISMA to recover the rMZR of the simulation by employing an observational approach. For this purpose, we considered three types of estimations, the intrinsic values, which are estimated directly from simulations; the predicted values, derived from the spectra generated by PRISMA; and observed values, calculated by using empirical and theoretical calibrations.

The paper is organized as follows. In §\ref{sec:simus} we describe the CIELO simulations and the simulated galaxy sample. In §\ref{sec:SPECGSG} we provide details about the pipeline and assumptions behind our numerical tool to create IFU-like datacubes of the simulated sample. Finally, in §\ref{sec:results} we used the synthetic spectra to estimate the star formation rate, analyze the performance of five metallicity indicators and calculate the rMZR.

\section{Simulations and Data selection}\label{sec:simus}

In this section, we provide details about the simulations used from the CIELO project and the selection criteria adopted to build the galaxy sample. This simulated sample was used as a test-bed for PRISMA; however, this code can be applied to any simulation that provides the required information.  

\subsection{The CIELO Project} 

CIELO stands for Chemo-dynamIcal propErties of gaLaxies and the cOsmic web, is a long-term project aimed at studying the formation of galaxies in different environments \citep{Tissera2025}. The CIELO runs were performed by using a version of GADGET-3, which includes treatments for metal-dependent radiative cooling, star formation, chemical enrichment and a feedback scheme for Type Ia and Type II supernovae  \citep{Scannapieco2005, Scannapieco2006}. The nucleosynthesis products from Type Ia and Type II supernovae were derived using the W7 model of \citet{Iwamoto1999} and the metallicity-dependent yields of \citet{Woosley_Weaver1995}, respectively.

 Cold and dense gas particles are transformed into stars according to temperature and density criteria \citep[e.g][]{pedrosatissera2015}. Each stellar particle represents a single stellar population (SSP) with the initial mass function (IMF) of \citet{Chabrier2003}. The CIELO simulations follow the evolution of 12 distinct chemical elements: He, C, Mg, O, Fe, Si, H, N, Ne, S, Ca, Zn,  where initially, baryons exist in the form of gas with primordial abundances $X_{\rm H} = 0.76$ and $Y_{\rm He} = 0.24$ \citep{Mosconi2001}.

The CIELO project encompasses a series of zoom-in simulations of different volume sampling low mass groups, filaments and walls. The simulations are consistent with $\Lambda$-Cold Dark Matter universe with $\Omega_0=0.317$, $\Omega_{\Lambda}=  0.6825$, $\Omega_B=0.049,$ h = $0.6711$  \citep{PlanckCollab2014}. In this paper we use the simulations named CIELO-LG1, CIELO-LG2,  CIELO-P3  and CIELO-P7 \citep{Tissera2025}. The three first simulations have dark matter and initial gas mass particles of $\rm m_{\rm dm} = 1.28\times 10^6~\rm M_\odot h^{-1}$ and $\rm m_{\rm gas} = 2.1 \times 10^5~\rm M_\odot h^{-1}$, respectively. CIELO-P7 has higher numerical resolution with $\rm m_{\rm dm}= 1.36\times 10^5 M_\odot h^{-1}$  and $\rm m_{\rm gas}= 2.1\times 10^4 M_\odot h^{-1}$.
The  gravitational softening for the intermediate (high) resolution simulations are $\sim 400 ~(250)$~pc  and $\sim 800 ~(500) $~kpc for baryons and dark matter (physical kpc at $z=0$), respectively.

We used the CIELO database \citep{Gonzalez-Jara2024}, where haloes are identified by applying a Friends-of-Friends (FoF) algorithm \citep{Davis1985} and the SUBFIND code selects the galaxies  within each of the dark matter haloes \citep{Springel2001,Dolag2009}. \citet{Tissera2025} reported the main parameters of the simulations and the fundamental relations they reproduced. CIELO galaxies satisfy the mass-size, the Tully-Fisher, the stellar mass-to-dark matter halo and the mass-metallicity relation. As discussed by these authors, these galaxies tend to be slightly less star forming than expected for the main sequence but they are still within observational bounderies. 

The CIELO galaxies have been previously used to study the impact of the environment on disc satellite  galaxies when entering a Local Group-like halo \citep{Rod2022}, the impact of baryon infall on the shape of central dark matter distribution \citep{Cataldi2023}, and the potential contribution of Primordial Black Holes to the dark matter component \citep{casanueva2024}.
 
\subsection{The CIELO galaxy sample} 

We select a sample of central CIELO galaxies with global characteristics similar to  galaxies in the MaNGA survey, since we intend to compare our results with observations from that project. It is important to clarify that the focus of this paper is not an in-depth analysis of the properties of these galaxies. Instead, we will use this sample to develop and validate our tool, with the intention of applying it to any simulated galaxy of interest. The following section details the criteria employed for this selection.

The MaNGA project is the largest IFU survey to date, which has observed more than 10000 galaxies in the local Universe at a redshift $z\sim 0.03$ \citep{Bundy2015}. This survey provides a sample of galaxies with stellar masses higher than $10^9$ $M_\odot$. Therefore, in order to have a simulated sample with similar stellar mass distribution, rescaled according to a \cite{Chabrier2003} IMF, we selected galaxies with $\rm \log(M_*/M_\odot) \ge 9$ at $z = 0$, where the stellar mass was computed considering all the stellar particles {\rm(selected by SUBFIND)} located within two times the optical radius \footnote{The optical radius is defined as the galactocentric radius that encloses $\sim 80$ per cent of the stellar mass of a galaxy.  
This  method allows the estimation of the spatial extension of a simulated galaxy \citep{tissera2010} and
 adapts to the distributions of each galaxy. Hence, it is preferable over using a fixed aperture \citep{Tissera2025}.}.

\begin{figure}[h!]
    \includegraphics[scale=0.65]{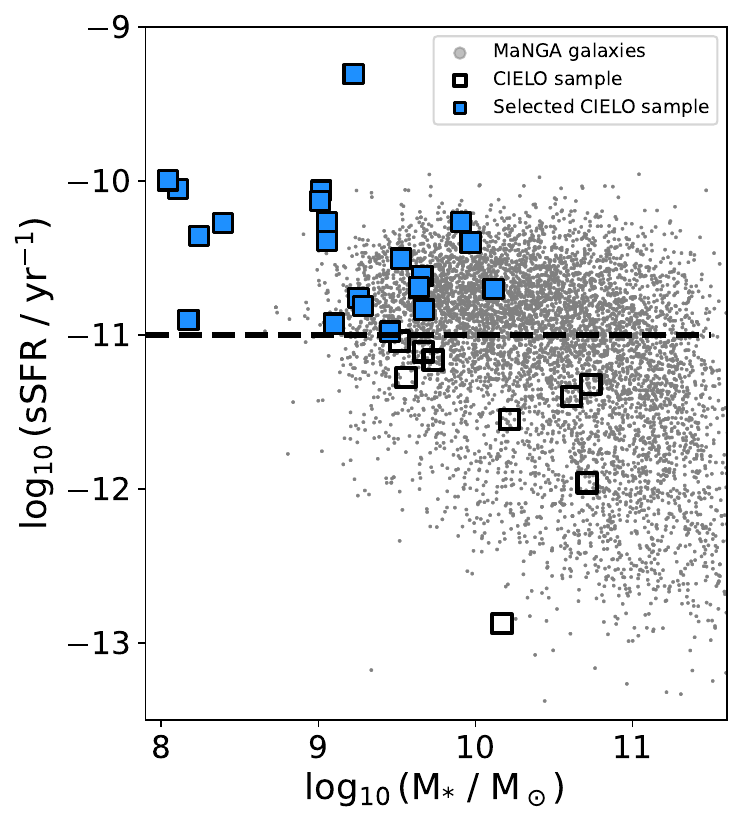}
    \caption{Specific star formation rate (sSFR) as a function of the stellar mass of galaxies from the CIELO project (squares) and the MaNGA survey (gray dots). Blue squares indicate the selected sample, which includes CIELO galaxies with $\rm log(M_\star/M_\odot) \geq 9$ and $\rm log(sSFR) > -11$ at $z = 0$. The unfilled black squares represent CIELO galaxies that were not included in the sample. The black dashed line marks the $\rm log(sSFR) = -11$.}
    \label{fig:hist_M_sSFR_Ropt}
\end{figure}

Figure~\ref{fig:hist_M_sSFR_Ropt} illustrates the relationship between the specific star formation rate (i.e., sSFR = SFR/M$_\ast$), and the stellar mass for this sample of CIELO galaxies (squares). Among these, we further selected those galaxies with a specific star formation rate, sSFR $> 10^{-11}$ yr$^{-1}$ \citep[e.g.,][]{Furlong2015}.

This threshold allows us to select star-forming galaxies whose resolved scaling relations have been widely studied with observations and simulations \citep[e.g.,][]{Sanchez2017, TrayfordSchaye2019}. This selection resulted in a sample of 21 star-forming galaxies (blue filled squares in Fig.~\ref{fig:hist_M_sSFR_Ropt}). 
For this level of star formation activity, the H$\alpha$ emission exceeds the minimum threshold established by \citet{Zhang2017} for minimizing contamination from Diffuse Ionized Gas (DIG). Specifically, they found that MaNGA spaxels with H$\alpha$ emission line surface density ($\rm \Sigma \text{H}\alpha = \text{H}\alpha / A_\text{spaxel}$, where $A_\text{spaxel}$ is the spaxel area) above $10^{39}$ erg s$^{-1}$ kpc$^{-2}$ show reduced DIG contamination in their spectra. Therefore, we expect minimal DIG impact on our sample.

For comparison, galaxies from the MaNGA survey, with $\rm log(M_\star/M_\odot) \gtrsim 9$ \citep[e.g.,][]{Ilbert2015, Spindler2018, Belfiore2018}, have also been included (gray dots). As can be seen, this sample of CIELO galaxies is within the observational range although they tend to show a lower sSFR than the median at a given stellar mass. We refer the reader to  \citet[][]{Tissera2025} for more details on the CIELO galaxies.

\section{PRISMA pipeline}\label{sec:SPECGSG}
The numerical tool PRISMA allows us to generate synthetic spectra of simulated galaxies, facilitating the determination of their physical properties from an observational perspective. The following sections provide details of the design of PRISMA as well as a description of the assumptions made for its construction. 

\subsection{Simulated Data Cubes}\label{sec:spaxs}

To accomplish this, we require resolved data cubes of the CIELO galaxies.  For this purpose, we construct grids that divide the projected mass distribution of each galaxy into hexagonal cells, hereafter spaxels, with an area $\rm s$. This scale is a free parameter within PRISMA and is, therefore, user-defined. For this work, we adopt $\rm s \sim 1$ \text{kpc}$^2$. The selection of this spaxel scale is motivated by the fact that we will confront the predicted rMZR with MaNGA results in the last section \citep{Barrera-Ballesteros2016}. However, the user can choose a different spaxel scale as needed.

Furthermore, PRISMA enables the 3D rotation of galaxies using a user-defined angle. This feature allows for inclination corrections, providing a valuable opportunity to explore both local and global properties of galaxies with different inclinations. For simplicity, in this work, we only analyzed face-on projections of the CIELO galaxies.

\begin{figure*}[h!]
        \centering
        \includegraphics[scale=0.32]{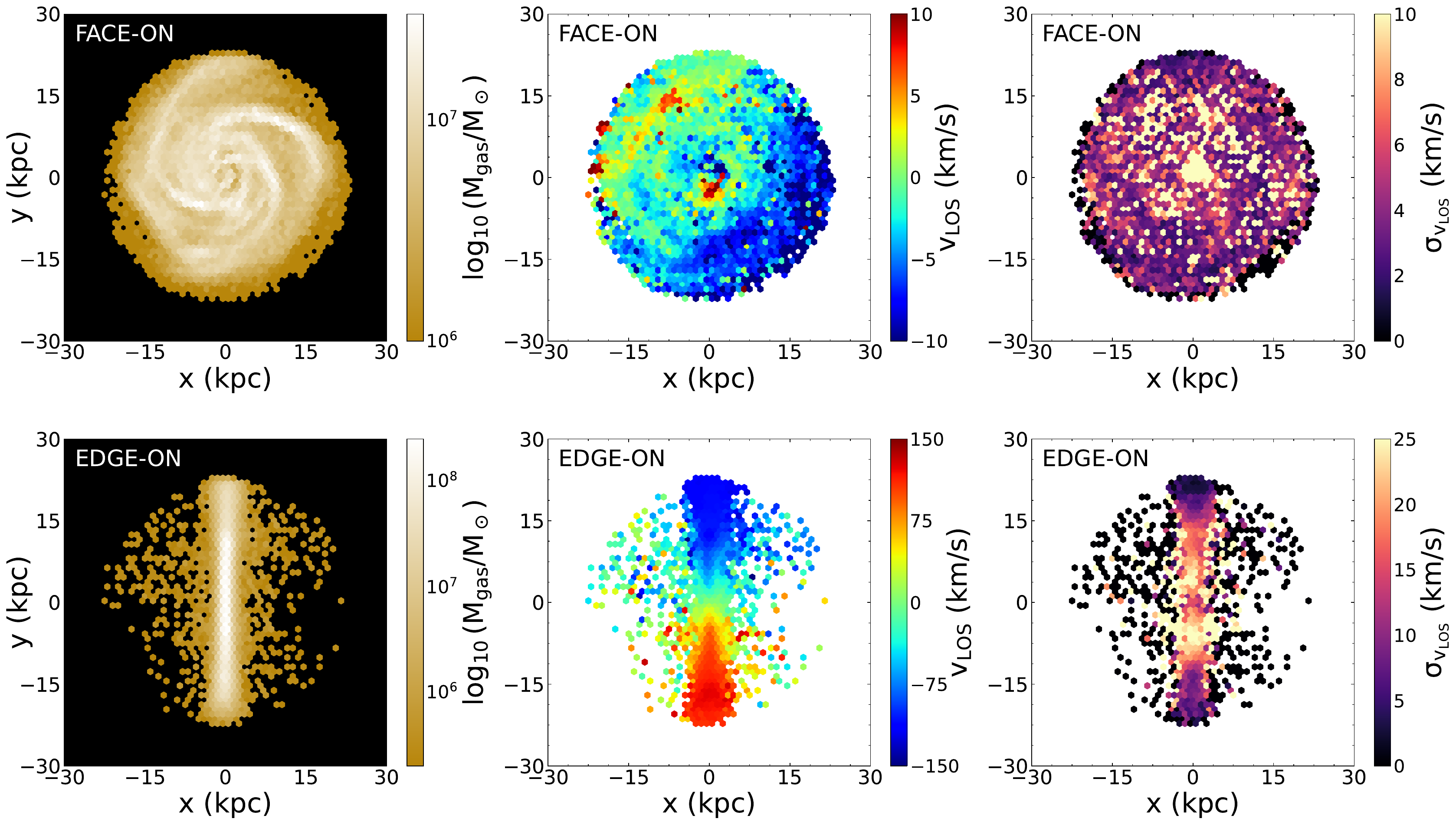}
        \caption{Face-on (top row) and edge-on (bottom row) projected properties of the gas component in a CIELO galaxy, as an example. All properties are estimated using the intrinsic properties from the CIELO database and adopting an spaxel scale of $s\sim 1$ kpc$^2$. The first column shows the gas-mass weighted distribution. The second column displays the velocity along the line of sight for the gas component, whereas the third column illustrates the dispersion of this velocity. The field of view is adjusted to $60 \times 60$~kpc$^2$ to encompass the entire galaxy. }
        \label{fig: gas grid}
\end{figure*}

Figure \ref{fig: gas grid} illustrates the two-dimensional properties of the gas distribution in a galaxy from the CIELO sample, considering its intrinsic information. The top row presents maps of the galaxy in a face-on orientation, while the bottom row displays maps in an edge-on orientation. The first column shows the spatial mass distribution of the gas component in the galaxy, while the second and third columns depict the line-of-sight (LOS) velocity and the velocity dispersion of the gas component, respectively (in km/s). The latter is calculated as the spread of the intrinsic velocities of the gas particles along the LOS within each spaxel, representing the deviation of the velocities from the mean velocity in that region. This figure reveals the morphology of the galaxy with a disk-like structure and discernible spiral arms, as can be seen in the upper panel of the first column. These features are consistent with the rotational patterns observed in the edge-on arrangement in the central column. Here, the expected red-shifted and blue-shifted velocity patterns of galaxies dominated by rotation are clearly detected.

\subsection{Synthetic spectra generation}\label{sec:spectra_generation} 

After generating the simulated data cubes, the next step is to compute the spectra emitted by the stellar populations and star-forming regions contained in each spaxel. For this purpose, we employ the Python software CIGALE (Code Investigating GALaxy Emission, \citep{Burgarella2005, Noll2009, Boquien2019}). This software facilitates spectrum generation and fitting by incorporating flexible models accounting for star formation history, stellar population emission, nebular emission, dust attenuation, among other properties. 

We utilized CIGALE to generate synthetic spectra emitted from individual spaxels. Within each spaxel, these spectra are computed by considering the contributions of the stellar populations according to their ages and metallicities as well as nebular emission, arising from the gas surrounding the star-forming regions. The subsequent sections describe the methodology used to calculate these spectra by integrating CIGALE with the information provided by the simulations.

\subsubsection{Star formation history} 

To generate the spectrum of each SSP, represented by a star particle, we assumed that its star formation history is quantified  by a star formation rate (SFR), which is characterized by a single starburst event. The SFR is calculated based on the age provided by the simulation for each SSP. Then, to model the SFH of each SSP, we employed the \texttt{sfhdelayed} scheme, which incorporates the expression SFR = $\rm \frac{t}{\tau^2} \times \exp(-t/\tau)$, where $\rm t$ represents the age of the stellar population and $\tau$ denotes the e-folding time. The e-folding time indicates the rate at which the SFR decays. In other words, a larger $\tau$ implies a more prolonged period of star formation, while a smaller $\tau$ indicates a rapid decline in star formation following the initial burst. In this work, we chose a small $\tau$ = 0.1 Myr to represent a quasi-instantaneous  birth of the entire stellar population.

\subsubsection{Stellar population emission} 

Having established the SFH model for each SSP, the next step is to compute their spectra. For this purpose, we adopt the version of the \cite{Bruzual-Charlot2003} stellar population synthesis models presented in \citet[][CB19 models hereafter]{Plat2019}. The stellar ingredients used in the CB19 models (referred to as C\&B models in some papers) are described in detail in \citet[][see their Appendix A]{Sanchez2022}. With PRISMA we can adopt a Salpeter or a Chabrier IMF. In alignment with the CIELO specifications, we adopt the Chabrier IMF, while Z and the age of each SSP are directly extracted from the simulated data. 

\subsubsection{Nebular emission}\label{sec:neb_emission}

When very massive stars form, they emit high-energy photons that ionize the surrounding gas. The ionized gas re-emits this energy in the form of emission lines and continuum, constituting what is known as nebular emission. This phenomenon depends on the chemical content and physical properties (electron density and temperature) of the interstellar medium and the ionization state of the gas. 

To account for the contribution of nebular emission to the stellar spectra, we follow a similar approach implemented in \citet{Nanni2022a}, where distinct procedures are employed based on the age of each SSP. On one hand, we designated SSPs younger than $10$ Myr as the young population, and the Z and t are taken directly from the simulations as mentioned above. The age threshold used in this study to separate between young and old SSPs is a user-defined parameter within our numerical tool. SSPs younger than this threshold are classified as the young population, and nebular emission is added to their spectra. SSPs older than this threshold are classified as the old population and do not include nebular emission in their spectra.

Additionally, we consider the star-forming (SF) gas as tracers of the regions of active star formation. SF gas particles are those that meet the conditions necessary for star formation (i.e., a temperature below $15,000$ K and a density exceeding $\rho_c = 7\times 10^{-25}$ g cm$^{-3}$, with $\rho_c$ the critical density), but because of the stochastic nature of star formation algorithm used, they have not been converted into star yet. In this case, we define fiducial young populations with fixed age of $10^5$ yr.  Properties such as mass, chemical abundances, density, position and velocities are inherited by the fiducial SSP from the progenitor SF-gas particle.
We assume that both types of particles represent the SF regions and serve as the primary contributors to generating nebular emission. However, the user can chose whether to include them.

To model the nebular emission, we follow different steps depending on whether the young population is represented by an individual stellar particle or an SF-gas particle. In the first case, the closest gas particle to each individual stellar particle is assumed to represent its surrounding ionized gas. We define the radius of the ionized gas cloud using the largest value between the smoothing length of the gas particle in question and the physical distance between it and the young SSP particle. This ensures that the population is contained within the gas clouds. 
In the second case, we assume that SF-gas particles represent their own gas clouds. In other words, we model these gas clouds using the properties of each SF-gas particle, including its smoothing length as the radius of the star-forming region, as well as its metallicity and density. 
Considering these two cases, when the young population is represented either by an individual stellar particle or by a SF-gas particle, we find that the median ionized gas cloud size in our sample with intermediate (high) numerical resolution is 530 (363) pc, with the $25^{\rm th}$ and $25^{\rm th}$ percentiles at 490 (353) pc and 560 (373) pc, respectively. 
These values are consistent with the range of giant star-forming region sizes observed in the literature, which typically span from 100 pc to several hundred parsecs, depending on the galaxy type and environmental conditions \citep[e.g.,][]{Hunt_Hirashita2009, KennicuttandEvans2012, Grasha2022}.

\begin{figure}[h!] 
\centering 
\includegraphics[scale=0.26]{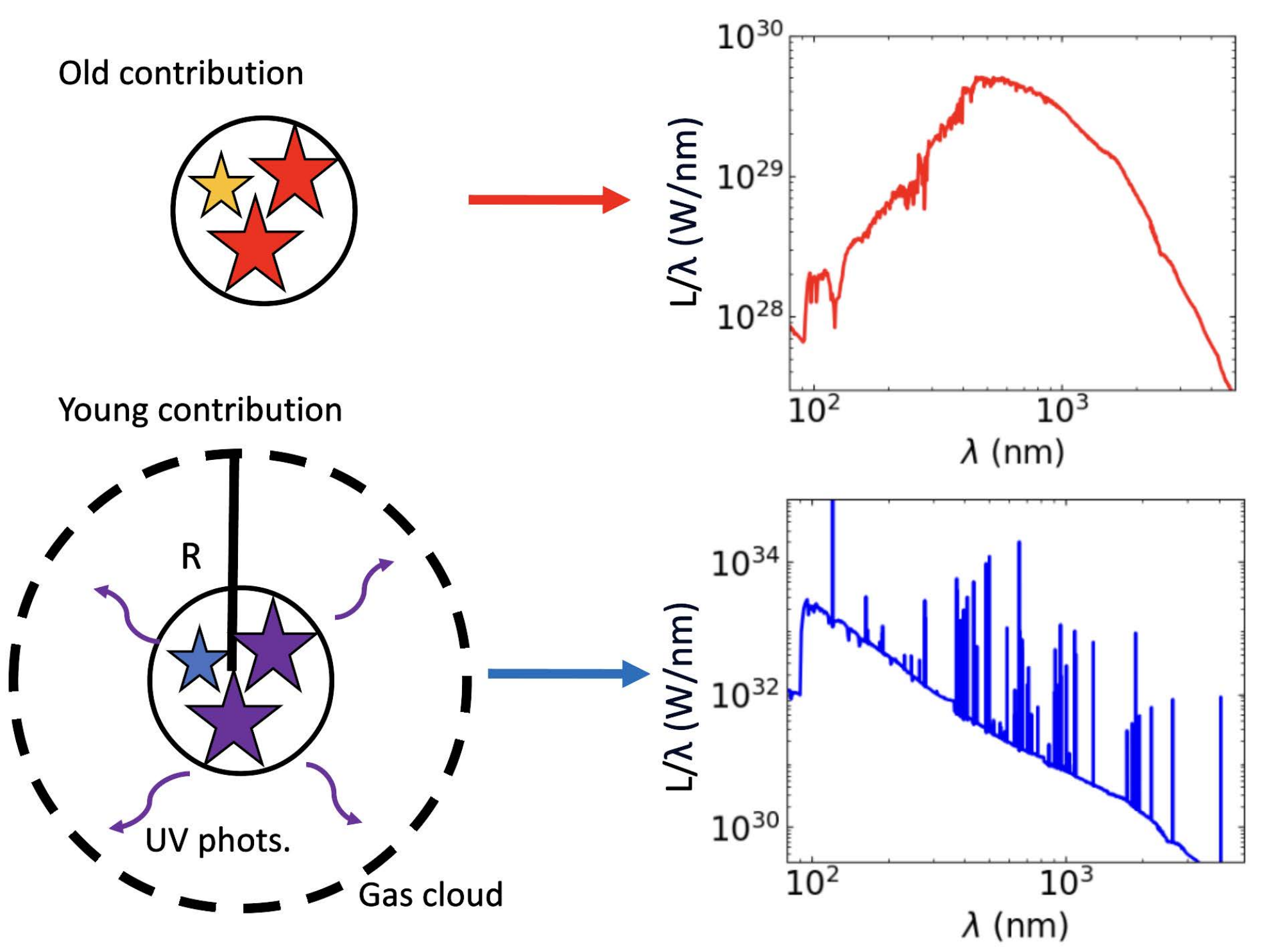} 
\caption{Diagram of the model we adopted to add the emission lines to the spectra in PRISMA. The inset top panel shows the spectrum produced by stellar populations older than $10$ Myr, which only depends on the CB19 stellar population synthesis models. The inset bottom panel displays the spectrum produced by stellar populations, which includes the CB19 SED models, and incorporates the nebular emission that is produced by the ionized gas cloud, which is represented by a sphere of radius R.} 
\label{fig: nebular model} 
\end{figure}

This assumption differentiates PRISMA from other tools generated in previous works \citep[e.g.,][]{Trayford2017, Rodriguez-Gomez2018, Schulz2020, Nanni2022a}. In these studies, the modeling of nebular emission considers direct properties of stellar populations (e.g., their metallicity) rather than the gas properties. Additionally, these works assumed fixed parameters of the ISM, such as its pressure and compactness\footnote{The compactness is a measure of the density of an H II region, which depends on the star cluster mass and the ISM pressure.}, which are intricately linked to the ionized state of the ISM. The hypothesis we adopt in our work allows us to model the nebular emission based on the current ionization state and the direct properties of the gas surrounding each simulated SSP. Figure~\ref{fig: nebular model} illustrates our model, presenting the spectrum generated by old SSPs (upper inset panel) and the spectrum generated when nebular emissions from young SSPs are considered (lower inset panel). 

CIGALE follows several steps to characterize nebular emission. Initially, leveraging the nebular templates proposed by \citet{Inoue2011}, it calculates the spectra of 124 emission lines as a function of $U$, the gas metallicity ($\rm Z_{\rm gas}$) and the electron number density ($n_{\rm e}$). Each emission line within each spectrum is shaped with a Gaussian profile, utilizing a user-defined line width. Additionally, because there is a possibility that the rate of ionizing photons is reduced due to photon escape from the medium or photon absorption by dust located between the emitting star and the surrounding gas, CIGALE provides the capability to reduce the intensity of nebular emission lines by considering both the escape factor ($\rm f_\text{esc}$) and the dust factor (f$_\text{dust}$). 

Hence, the free parameters employed by CIGALE to describe the nebular emission are $\rm U$, Z$_\text{gas}$, $\rm n_{\rm e}$, the line width, $\rm f_\text{esc}$, and f$_\text{dust}$. In this context, we determine each of these parameters individually for each young SSP. Below we elaborate on how each parameter is defined based on the characteristics of the stellar populations and the surrounding gas. 

\paragraph{Ionization parameter:}We compute the ionization parameter U following the equation \ref{eq: logU}, where $\rm R$ denotes the radius of the ionized gas cloud, $\rm f_\text{esc}$ represents the escape fraction of photons that do not contribute to the ionization of the gas cloud, f$_\text{dust}$ corresponds to the fraction of photons that are absorbed by dust within the star-forming region, $\rm n_{HI}$ is the numerical density of neutral hydrogen, $\rm c$ stands for the speed of light, and $\rm Q_\text{ion}$ represents the rate of photons ionizing the gas cloud.  

    \begin{equation} 
    \centering 
    \rm U = \frac{\rm Q_\text{ion}(1-\rm f_\text{esc}-\rm f_\text{dust})}{\rm 4\pi R^2 n_{HI} c} 
    \label{eq: logU}  
    \end{equation} 

    To calculate $\rm Q_\text{ion}$, we consider both the rate of ionizing photons emitted by each stellar population ($\rm Q_*$) and the maximum rate of ionizing photons that can be captured by the surrounding gas to become completely ionized ($\rm Q_\text{gas}$). On the one hand, $\rm Q_*$ is calculated from the SSP models of CB19 that depends on the age, mass and metallicity of the stellar populations. On the other hand, $\rm Q_\text{gas}$ is calculated by considering a simplified model where the gas cloud is composed only by hydrogen atoms, with a recombination rate $\rm \alpha_H\sim 2\times 10^{-13} \text{cm}^3 \text{s}^{-1}$ \citep{Ferland1980} and an electron number density of $\rm n _e$. With that, the maximum ionizing photon rate that can be received by each gas particle is $\rm Q_\text{gas}=\alpha_H n_e^2 4\pi R^3/3$.  

    Therefore, if $\rm Q_* > Q_\text{gas}$, the gas cloud is considered to be completely ionized by the radiation emitted by the stellar population, and hence we have adopted, $\rm Q_\text{ion}= Q_\text{gas}$. Conversely, if $\rm Q_* < Q_\text{gas}$, then the gas cloud is considered to be partially ionized, with $\rm Q_\text{ion}= Q_*$. In our sample, for all stellar populations we find that $\rm Q_*> Q_\text{gas}$, so all star-forming regions in the sample are fully ionized. 

\paragraph{Gas metallicity:}We considered the metallicity of the gas cloud, which is directly provided by the simulation. 

\paragraph{Electron number density:} 
    The simulation directly provides the initial electron number density ($\rm n_e$) of the gas particles surrounding each SSP. However, this initial $\rm n_e$ does not account for the ionization impact of young stars.
    
    When the gas cloud undergoes ionization, its electron density will vary in response to the incoming ionizing photons. As we specified earlier, when $\rm Q_* > Q_\text{gas}$, the gas cloud is assumed to be completely ionized. In this case, considering that $\rm n_{HI}$ is the initial neutral hydrogen density directly provided by the simulation, the resulting electron number density is given by $\rm \widetilde{n_e} = n_e + n_{HI}$. In contrast, when $\rm Q_* < Q_\text{gas}$, $\rm \widetilde{n_e}$ is determined based on the hydrogen atoms that have been ionized, considering the rate of ionizing photons emitted by the stellar population. Because  $\rm Q_* > Q_\text{gas}$ for all stellar populations in our sample, we adopt $\rm \widetilde{n_e} = n_e + n_{HI}$, where both $\rm n_e$ and $\rm n_{HI}$ are quantities directly provided by the simulation.
    
    To use CIGALE, $\rm n_e$ is required to be binned between $[10, 100, 1000]$ cm $^{-3}$. Consequently, we assigned the simulated $\rm \widetilde{n_e}$ values to  the closest value specified by CIGALE. We investigate the impact of using  $\rm n_e=[10, 100, 1000]$ cm$^{-3}$ on our calculations. Our findings revealed that these variations do not significantly alter the computed values of SFR and chemical abundances.  In particular, we estimated a variation of 0.002 dex in the median SFR and a maximum variation of 0.06 dex in the median of the chemical abundances when using $\rm n_e=[100]$ and $\rm n_e=[1000]$ with respect to $\rm n_e=[10]$, respectively. This latter estimation was obtained considering the five abundance indicators used in this paper (see Sect. \ref{sec:determining_abundances}).
    
\paragraph{Line width:} The line width in the spectra reflects the kinematics of the gas within each region. The higher the velocity dispersion along the LOS, the broader the linewidth in the spectrum.  To define these widths, we considered the standard deviation of the velocities along the LOS of all the gas particles that are located within each spaxel (see third column of Fig.~\ref{fig: gas grid}). 

\paragraph{f$_\text{esc}$ and f$_\text{dust}$:} Both parameters describe the fraction of photons that do not contribute to the nebular emission because they either escape ($\rm f_\text{esc}$) or are absorbed by dust within the SF region (f$_\text{dust}$). We set the $\rm f_\text{esc}= 0.10$, which implies that 10\% of the photons escape and do not ionize the gas cloud. This assumption is supported by observations at low redshift indicating that the escape fraction of Lyman-continuum photons does not exceed 10\% \citep[e.g.,][]{Bergvall2006, Plat2019}, and references therein). To calculate f$_\text{dust}$, we consider this factor as a free parameter, adjusting it to the value that produces spectra enabling the best recovery of the intrinsic SFR. This parameter allows us to incorporate a simple model to take into account the effect of  dust inside the SF region. This dust is assumed to form a cocoon  which absorbs a fraction of  UV photons emitted by the young stellar population.

\subsubsection{Dust attenuation} 

Dust within the ISM can modify galaxy spectra through processes such as absorption, scattering, and re-emission of photons, along the LOS. Dust absorbs UV to near-infrared photons and subsequently re-emits them in the mid- and far-infrared wavelengths. This principle of energy balance lies at the core of CIGALE, which allows the convolution of the spectra with the Calzetti attenuation law \citep{Calzetti2000}.

We investigate the effects of dust attenuation to assess how they could affect our calculations in Sect. \ref{sec:application_of_dust_att}. For this purpose, we created two datasets of spectra from the same sample of spaxels, considering different scenarios of dust absorption. The first dataset consists of spectra that are not attenuated by dust along the LOS. Therefore, they are only affected by dust within the \ion{H}{ii} regions, which absorbs the most energetic photons (regulated by the f$_\text{dust}$ parameter). The second dataset comprises spectra that are attenuated by both the dust within the \ion{H}{ii} regions and the dust in the ISM along the LOS. To model the impact of the latter, we used the Calzetti attenuation law by adopting $E(B-V) = 0.3$ mag. This approach provides a simplified representation compared to full radiative transfer treatments \citep[e.g.,][]{Barrientos_Acevedo2023}. However, the analysis by \citet{Nanni2022a} suggest that the effects of employing the Calzetti law are comparable to those obtained by using the Monte Carlo radiative transfer code SKIRT \citep{Baes2011, Baes_Camps2015}.

\section{Results and Discussion}\label{sec:results}

In this section, we apply PRISMA to the CIELO galaxies to compute the synthetic spectra with different dust recipes and use five metallicity indicators to estimate predicted abundances.  

\subsection{Fitting the free parameters of PRISMA}

The first step is to choose the free parameters of PRISMA and to do this we request PRISMA to recover the intrinsic SFR from the H$\alpha$ emission line flux.

\subsubsection{Recovering the intrinsic SFR without dust attenuation}\label{cap: param. setting} 

To calculate the intrinsic SFR per spaxel, we consider the total mass of the  SSPs  younger than $\rm t = 10^7$ yrs in a given spaxel, ${\rm SFR= M_{young}/t}$. Then, we determine the predicted SFR per spaxel by summing the synthetic H$_\alpha$ emission line flux generated by the SSPs within a given spaxel. Subsequently, we converted this H$_\alpha$ flux into the predicted SFR by using the calibration reported by \citet{Calzetti2013}: 

\begin{equation}\label{eq: SFR} 
    \rm \text{SFR}\left(\frac{\text{M}_\odot}{\text{yr}}\right) = 5.5 \times 10 ^{-42} \text{L(H}_\alpha) \left(\frac{\text{erg}}{\text{s}}\right)  
\end{equation} 

This equation is suitable for a Chabrier IMF \citep{KennicuttandEvans2012}, therefore, it aligns well with the CIELO specifications. Moreover, a comparable version of this relation was employed by \citet{Baker2023} to estimate the SFR in spatially resolved regions of MaNGA galaxies.

As previously mentioned, CIGALE utilizes various free parameters for characterizing nebular emission. Most of these parameters are derived from the intrinsic data of the simulation (e.g., log U, $\rm Z_\text{gas}$, $\rm n_e$, and the line width) as described in previous sections. However, the $\rm f_\text{esc}$ and  f$_\text{dust}$  need to be determined. Since we are working with galaxies at $\rm z=0$, as we mentioned above, we assume  $\rm f_\text{esc}=0.1$. For the f$_\text{dust}$, we take three different values: f$_\text{dust}=0.4, 0.5,$ and $0.6$  to select the one that best traces the intrinsic SFR of the simulated sample. For all these three cases, we consider no attenuation by dust along the LOS.

\begin{figure}
    \includegraphics[scale=0.5]{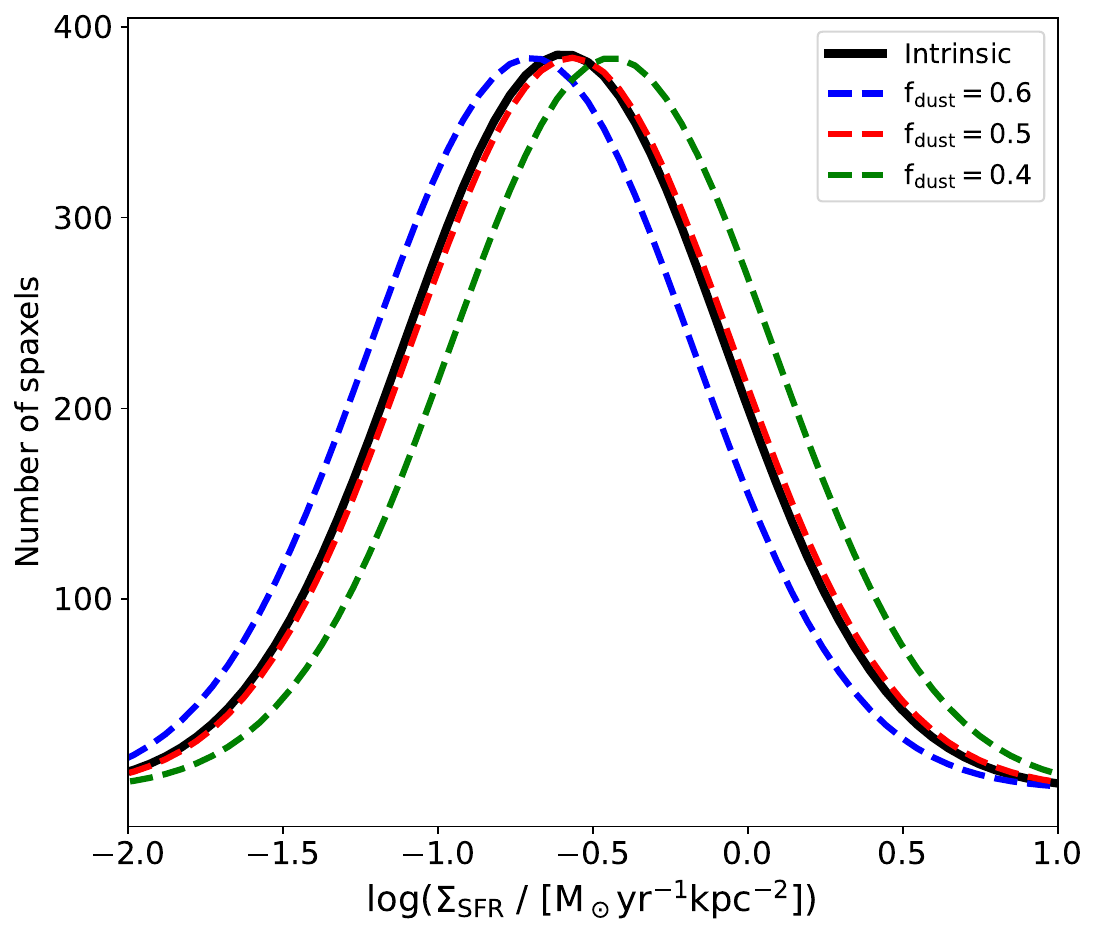}
    \caption{Gaussian distributions fitted to the intrinsic SFR surface densities calculated by spaxel (black, solid lines) and to the corresponding predicted distributions obtained by applying PRISMA with  f$_\text{dust} = 0.4$, 0.5, and $ 0.6$ (green, red and blue, dashed lines, respectively.}    
    \label{fig:setting_fdust}
\end{figure}

Hence, we applied PRISMA with different f$_\text{dust}$ and estimated the predicted SFR and the SFR surface densities in each case, defined as $\Sigma_{\rm SFR} = \rm SFR/A_{\rm spaxel}$ where $\rm A_{spaxel}$ is the area of the spaxel. Then, we built histograms of $\Sigma_{\rm SFR}$ and fit them by Gaussian distributions to better compared the results obtained from different f$_\text{dust}$. Figure~\ref{fig:setting_fdust} shows the comparison of the Gaussian fittings to the intrinsic (black, solid line) and the predicted (dashed lines) $\Sigma_{\rm SFR}$ for  runs with f$_\text{dust} = 0.4, 0.5, 0.6$ (green, red and blue, respectively). From this panel, we can see that the predicted distribution obtained with f$_\text{dust}=0.5$ is the one that best recovers the intrinsic SFR distribution.

Additionally, to achieve a better quantification, we calculated the relative difference between the intrinsic and predicted SFR. We define it as $\rm \Delta(SFR) = (SFR_{predicted} - SFR_{intrinsic}) ~ / ~ SFR_{intrinsic}$. For f$_\text{dust}=0.5$, which is the parameter that better reproduces the intrinsic values, we obtain a median $\rm \Delta(SFR) = 0.06$, with the $25^{\rm th}$ and $75^{\rm th}$ percentiles at 0.04 and 0.07, respectively. This indicates that PRISMA slightly overestimates the predicted SFR values with a relative error of 6\%. Therefore, hereafter, we use f$_\text{dust}=0.5$ to generate the synthetic spectra of the CIELO sample, from which  their resolved gas-phase metallicities will be derived for comparison with MaNGA observations.

\begin{figure*}[h!]
    \includegraphics[scale=0.7]{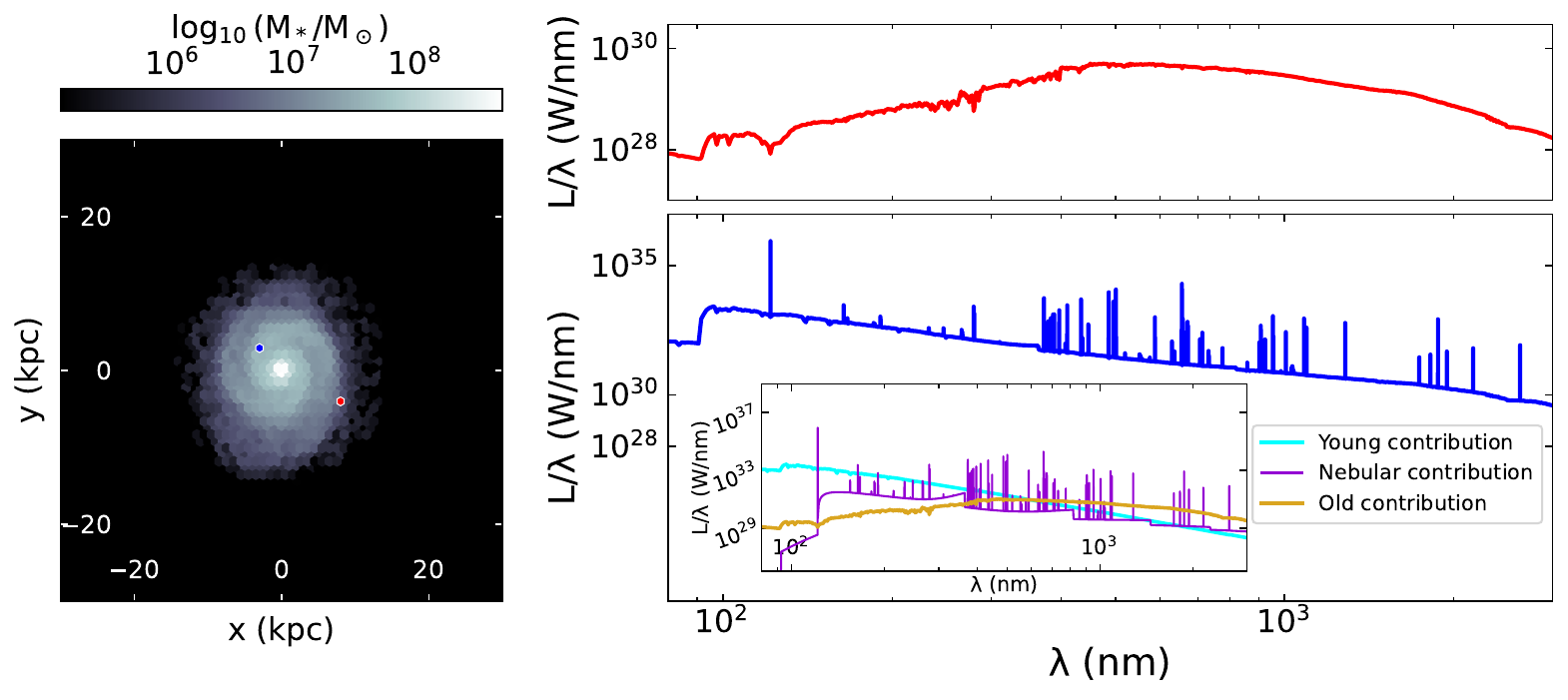}
        \caption{Example of the composite spectra generated with PRISMA for two spaxels in a face-on galaxy of our CIELO sample. Left panel: the stellar mass-weighted distribution of a given galaxy, where two spaxels  are highlighted in red and blue. The red spaxel contains only stellar populations older than 10 Myr, whereas the blue spaxel contains both old and younger stellar populations. Right panels: the corresponding spectra obtained with PRISMA from each of these spaxels (red and blue spectra, respectively). Additionally, the inset plot shows the old, young and nebular contributions (orange, cyan and violet, respectively) that  compose the  total spectrum of the blue spaxel. These spectra were not attenuated by ISM dust along the LOS.}
    \label{fig:mocked_spectra}
\end{figure*}

Figure~\ref{fig:mocked_spectra} illustrates examples of spectra computed by PRISMA (excluding dust attenuation along the line of sight) for a face-on galaxy from the CIELO sample (which is different from the galaxy shown in Fig.~\ref{fig: gas grid}). The left map displays the stellar mass distribution of the galaxy, with two highlighted spaxels, representing different regions. The red spaxel is located in an outer region of the disc, exclusively containing stellar populations older than $10^7$ yr. Therefore, its composite spectrum has no emission lines (red spectrum). The blue spaxel is situated in one of the galaxy’s spiral arms, containing both young and old stellar populations. The spectra produced by each of these populations are shown in the inset of the lower right panel of Fig.~\ref{fig:mocked_spectra} in cyan and orange, respectively. Additionally, the  contribution of the nebular emissions from the star-forming zones that enclose the young stars is  displayed separately (violet spectrum). By accumulating all these contributions, we obtained the total spectrum emitted in this region as shown by the blue spectrum.

\subsubsection{Recovering the intrinsic SFR with dust attenuation}\label{sec:application_of_dust_att}

Until now, our discussion has not considered the absorption and re-emission effects of the computed spectra by dust  that could be present in the ISM along the line of sight. However, as discussed in Sect. \ref{sec:neb_emission}, we have included a simple model for dust absorption within star-forming regions adopting the parameter f$_\text{dust}$ of CIGALE. In order to assess the impact of the ISM dust on our estimations, we have additionally produced spectra attenuated by the Calzetti law, following the procedure explained in Sect.~\ref{sec:spectra_generation}. 

To generate these attenuated spectra, the f$_\text{dust}$ value in PRISMA was taken as a free parameter in order to recover the intrinsic SFR from the attenuated spectra, similarly to was explained in the previous section. By reducing f$_\text{dust}$ from $0.5$  to $0.1$, we succeeded in recovering the intrinsic SFR of the sample, where we registered a change of $\rm \Delta SFR = -0.005^{+0.005}_{-0.007}$. With this value of f$_\text{dust}$, only 10\% of the photons are now absorbed by the dust within the star-forming regions while the rest of the attenuation effects are associated with the impact of dust along the line of sight.

\subsection{Estimating the gas-phase metallicity}\label{sec:determining_abundances}

Considering $\rm f_\text{esc}=0.1$, f$_\text{dust}=0.5$ for the non-attenuated case and $\rm f_\text{esc}=0.1$, f$_\text{dust}=0.1$ for the case run with the Calzetti law, we generated synthetic spectra with emission lines from the simulated sample. To estimate the intrinsic gas-phase metallicity for each spaxel, we used the intrinsic oxygen and hydrogen abundances provided directly by the simulation. Specifically, for each spaxel, we summed the total oxygen and hydrogen contributions from the gas particles assumed to represent star-forming regions. This total was then expressed in units of 12 + log(O/H). We focus exclusively on these regions because we later compare these intrinsic values with the estimates derived from their emission.

Furthermore, we derived the predicted gas-phase metallicity by employing emission line diagnostics that establish a connection between metallicity and emission line ratios originating from different chemical elements. Specifically, we utilized the N2, R23, $\hat{\rm R}$, O3N2 and N2O2 indicators, which are defined in Table~\ref{tab: Indicators_definition}. These indicators and their calibrations were chosen because they have been widely applied in observational studies \citep[e.g.,][among others]{Pettini_Pagel2004, Maiolino2008, PerezMontero_Contini2009, Marino2013, Zhang2017, Curti2017, Curti2020, Laseter2023}. Additionally, they have been previously employed in simulation-based studies that generate synthetic spectra with emission lines, such as \cite{Garg2023} for SIMBA and \cite{Hirschmann2023} for IllustrisTNG simulations. A detailed discussion on them can be found in \citet{Kewley_Ellison2008} and \citet{Conroy2013}, for example. We stress the fact that three different estimations for the SFR and the metallicity indicators will be considered: the intrinsic values, which are directly estimated from the simulations, the predicted values, given by PRISMA and, the observed values, calculated by using empirical and theoretical calibrations. 

\begin{figure} 
    +\hspace*{-0.7cm}  
    \includegraphics[scale=0.53]{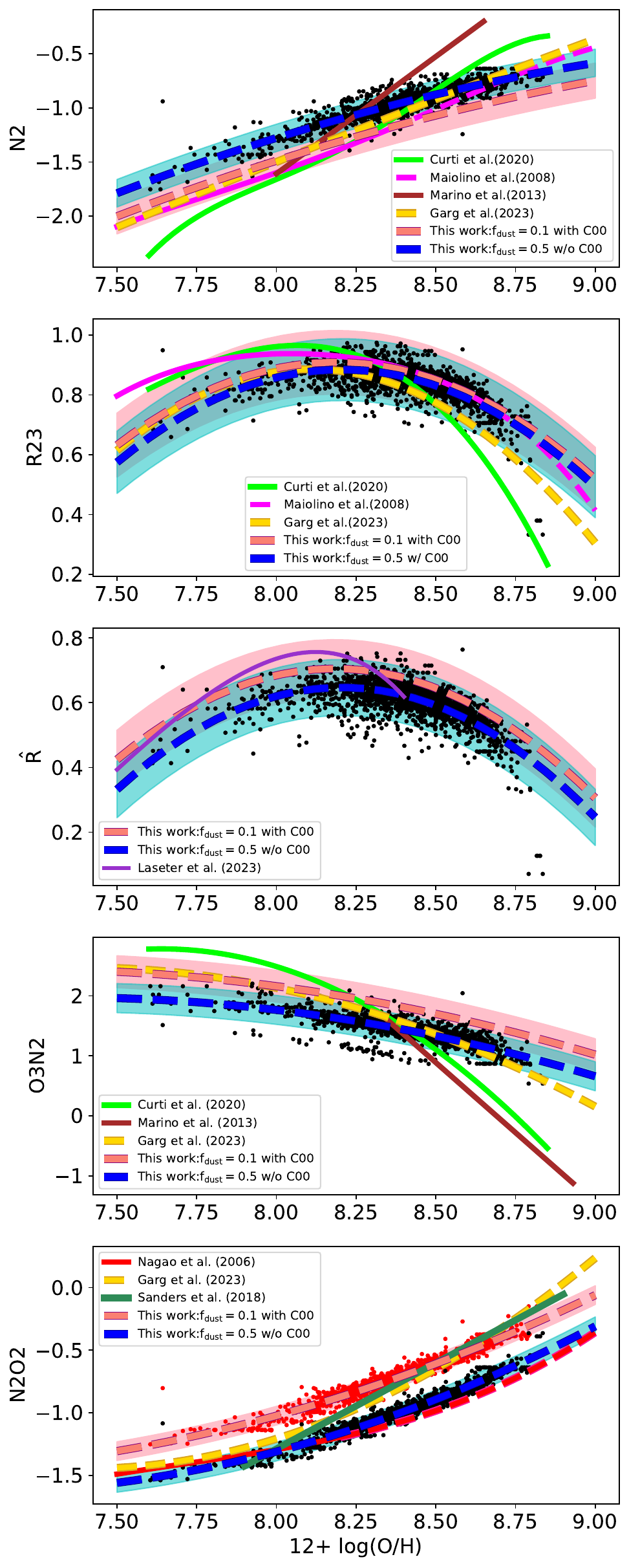} 
    \caption{Emission line ratios as a function of the intrinsic oxygen abundance for the sample data (black dots). The blue and pink dashed lines represent the best fits to the simulated data, considering synthetic spectra non-attenuated and attenuated by the Calzetti law, respectively. The shaded regions shows the 95 percent prediction intervals. The red dots in the last panel represent the attenuated data by the Calzetti law. %confidence intervals. 
    The metallicity range of applicability of each literature calibration is reflected in the length of each relation.}
    \label{fig:calibrators} 
\end{figure}

We first analyze the non-attenuated case. Figure~\ref{fig:calibrators} shows the predicted N2, R23, $\hat{\rm R}$, O3N2 and O2N2 ratios  derived from synthetic CIELO spectra as function of intrinsic $12 + \log(\rm O/H) $ (black filled circles). We fit these data using a polynomial function of the form $\rm y = \rm c_0 + \rm c_1 x + \rm c_2 x^2$, where $\rm x = 12$ + log(O/H) and $\rm y$ is the respective line ratio. The fit to data lacking dust attenuation along the LOS is represented by the blue dashed lines, while the pink dashed curves show the fit to the data that were influenced by dust along the LOS. The shaded regions correspond to the 95 per cent prediction intervals, which account for the expected dispersion of individual data points around the fitted polynomial. Table \ref{tab: indicators_fit_parameters} displays the coefficients of the polynomial fits for all the metallicity indicators discussed above, for the spectra lacking dust attenuation along the LOS. This table also includes the lower and upper limits of the 95\% prediction interval. Similar fits were performed for the attenuated case (see Table~\ref{tab: indicators_fit_parameters_clz} for more details). Additionally, for better comparison, the red dots in the last panel of the figure show the distribution of the data attenuated using the Calzetti law.

Figure \ref{fig:calibrators} also includes the empirical calibrations derived by \citet{Marino2013, Curti2020, Sanders2017, Laseter2023}, shown by the brown, light green, dark green, and purple solid lines, respectively. For comparison, the calibration reported by \citet[][yellow, dashed line]{Garg2023} is also shown, which was obtained using their own mocked emission method applied to the cosmological simulation SIMBA  at $z=0$ \citep{Dave2019}. Their predicted values were derived with the photoionization code Cloudy \citep{Ferland2017} and considering the contributions from \ion{H}{ii} regions, post-AGB stars and DIGs. The relations provided by \citet{Maiolino2008} and \citet{Nagao2006} are semi-empirical calibrations (magenta and red lines, respectively). They used photoionization models to extend their empirical calibrations in the regime where there is no observational data available. 

As can be seen, the N2 and N2O2 curves, which directly depend on \text{[\ion{N}{ii}]}$\lambda6584$, show a significant increase with increasing oxygen abundance, which is mainly a consequence of secondary nitrogen\footnote{Secondary nitrogen is produced from CNO elements present in stars, so that the nitrogen yield is metallicity-dependent \citep{Clayton_1983}} production towards higher metallicities. The $R23$ and $\hat{R}$ ratios exhibit a more complex dependence on oxygen abundance because they show double-valued functions of metallicity. For metallicities below 12 + log(O/H) $\sim 8.2$, these ratios increase for increasing O/H due to growing oxygen abundance. However, for higher metallicities, these ratios decrease as oxygen abundance rises. This is a result of efficient cooling of gas by heavy elements through metal lines, leading to fewer collisional excitations of optical transitions. Finally, the O3N2 ratio demonstrates a decreasing trend, aligning with the aforementioned arguments. As 12 + log(O/H) becomes higher, both the increase in [\ion{N}{ii}]$\lambda6584$ and the cooling of the gas cause the line ratio [\ion{O}{iii}]$\lambda5007 ~ / ~ [\text{\ion{N}{ii}}] \lambda6584$ (i.e., O3N2) to decrease.

\begin{table}[] 
\caption{Line ratios definition} 
\centering 
    \begin{tabular}{lcc}
    \hline \hline
    Notation &  Line Ratio \\ 
    \hline 
    N2              & $\rm log(\text{[\ion{N}{ii}]}_{\lambda6583} ~/~ \rm H_{\alpha})$ \\ 
    R32             & $\rm log((\text{[\ion{O}{ii}]}_{\lambda3726,29} + \text{[\ion{O}{iii}]}_{\lambda4959,5007})~/~\rm H_{\beta})$ \\
    $\hat{\rm R}$         & $0.47 \times \rm R2 + 0.88 \times \rm R3$ \\
    O3N2            & $\rm log((\text{[\ion{O}{iii}]}_{\lambda5007}/\rm H_{\beta})~/~(\text{[\ion{N}{ii}]}_{\lambda6583}/\rm H_{\alpha}))$ \\
    N2O2            & $\rm log(\text{[\ion{N}{ii}]}_{\lambda6583} ~/~ \text{[\ion{O}{ii}]}_{\lambda3726,29})$ \\
    \hline
    \end{tabular} 
\label{tab: Indicators_definition} 
\end{table} 

\begin{table}[]
\caption{Predicted metallicity calibrations.}\label{tab: indicators_fit_parameters}
\centering
\begin{tabular}{lllll}
\hline \hline
Line ratio      &     Fit  & Lower curve &  Upper curve
\\ \hline
%N2
\begin{tabular}[c]{@{}c@{}}N2\\ \end{tabular} & \begin{tabular}[]{@{}l@{}}$\rm c_0=-21.818$\\ $\rm c_1= 4.234$\\ $\rm c_2 = -0.208$\end{tabular}   & \begin{tabular}[]{@{}l@{}}$\rm c_0= -22.016$\\ $\rm c_1= 4.250$\\ $\rm c_2 = -0.209$\end{tabular} & \begin{tabular}[]{@{}l@{}}$\rm c_0= -21.621$\\ $\rm c_1= 4.217$\\ $\rm c_2 =  -0.207$\end{tabular}  \\ \hline
%R23
\begin{tabular}[c]{@{}c@{}}R23\end{tabular}& \begin{tabular}[]{@{}l@{}}$\rm c_0= -41.467$\\ $\rm c_1= 10.320$\\ $\rm c_2= -0.629$\end{tabular}  & \begin{tabular}[]{@{}l@{}}$\rm c_0= -41.629$\\ $\rm c_1= 10.333$\\ $\rm c_2= -0.629$\end{tabular} & \begin{tabular}[]{@{}l@{}}$\rm c_0= -41.305$\\ $\rm c_1= 10.306$\\ $\rm c_2= -0.628$\end{tabular}   \\ \hline
%Rhat
\begin{tabular}[c]{@{}c@{}}$\hat{\rm R}$\\\end{tabular}& \begin{tabular}[]{@{}l@{}}$\rm c_0= -42.769$\\ $\rm c_1 = 10.580$\\ $\rm c_2= -0.645$\end{tabular} & \begin{tabular}[]{@{}l@{}}$\rm c_0= -42.905$\\ $\rm c_1= 10.591$\\ $\rm c_2= -0.645$\end{tabular} & \begin{tabular}[]{@{}l@{}}$\rm c_0= -42.633$\\ $\rm c_1= 10.568$\\ $\rm c_2= -0.644$\end{tabular}   \\ \hline
%O3N2
\begin{tabular}[c]{@{}c@{}}O3N2\\ \end{tabular}& \begin{tabular}[]{@{}l@{}}$\rm c_0= -26.193$\\ $\rm c_1= 7.590$\\ $\rm c_2= -0.512$\end{tabular}   & \begin{tabular}[]{@{}l@{}}$\rm c_0= -26.573$\\ $\rm c_1= 7.621$\\ $\rm c_2= -0.514$\end{tabular} & \begin{tabular}[]{@{}l@{}}$\rm c_0 = -25.813$\\ $\rm c_1 = 7.558$\\ $\rm c_2= -0.510$\end{tabular} \\ \hline
%N2O2
\begin{tabular}[c]{@{}c@{}}N2O2\\ \end{tabular}& \begin{tabular}[]{@{}l@{}}$\rm c_0= 14.810$\\ $\rm c_1= -4.695$\\ $\rm c_2= 0.335$\end{tabular}    & \begin{tabular}[]{@{}l@{}}$\rm c_0= 14.694$\\ $\rm c_1= -4.685$\\ $\rm c_2= 0.334$\end{tabular}   & \begin{tabular}[]{@{}l@{}}$\rm c_0= 14.926$\\ $\rm c_1= -4.705$\\ $\rm c_2= 0.336$\end{tabular}     \\ \hline
\end{tabular}
\tablefoot{Best fitting relations for the synthetic line ratios displayed in Fig. \ref{fig:calibrators} at $z=0$, considering the spectra non-attenuated by ISM dust along the LOS (blue lines). The second column shows the fit parameters of our calibration (Sect. \ref{sec:determining_abundances}), while the third and fourth columns show the fit parameters for the lower and upper 95\% confidence intervals.}
\end{table}

Overall, there is agreement between published calibrations and those predicted by our model for CIELO galaxies. However, the level of agreement varies depending on the calibrator used. Some differences exist for the O3N2 calibrations at $\rm 12+ {log(O/H)} > 8.5$, where the simulated ratios are higher than expected from the empirical relations of \citet{Marino2013} and \citet{Curti2020}, but agree better with the numerical predictions of \citet{Garg2023}. A similar trend is observed for the R23 indicator at 12 + log(O/H) > 8.5, where our fit is in good agreement with the numerical predictions of \citet{Garg2023} and the semi-empirical calibration of \citet{Maiolino2008}, but our slope differs from the calibration of \citet{Curti2020}. 

The differences between calibrations based on electron temperature \citep[e.g.,][]{Curti2020} and those derived from photoionization models (e.g., \citealt{Garg2023} and \citealt{Maiolino2008} in the high metallicity regime) are well-documented in the literature \citep[e.g.,][]{Stasinska2005, Kewley2008, Cameron2023}. At high metallicities, the $\rm T_{\rm e}$ method tends to saturate and significantly underestimate the true metallicity. This difference arises from temperature fluctuations and gradients, both within individual \ion{H}{ii} regions and across the entire galaxy \citep{Stasinska2005, Cameron2023}.

Regarding our differences with \citet{Garg2023}, we note that they do not consider Active Galactic Nucleus (AGN) emission in their analysis, which makes it unlikely that AGNs contribute to the observed discrepancies, since our simulations neither include this process. However, their study does include the contribution from DIG, post-AGB stars, and incorporates dust in their simulations. DIG contamination alone may not fully account for the discrepancy as we only considered spaxels with $\rm \Sigma H_\alpha > 10^{39} \text{erg s}^{-1} \text{kpc}^{-2}$, which are less affected according to \citet{Zhang2017}. Additionally, for the N2O2 index as an example, our predictions closely match the line ratios presented in \citet{Nagao2006}, but deviate from the calibrations of \citet{Sanders2018} and \citet{Garg2023} for $\rm 12+log(O/H) > 8.15$, where these calibrations have a steeper slope. The difference between the two observational calibrations can be explained by the fact that \citet{Sanders2018} applied a correction for DIG contamination whereas \citet{Nagao2006} did not. However, \citet{Garg2023} still obtained a slope similar to that of \citet{Sanders2018}. This reinforces the idea  that DIG contamination may not necessarily be the factor causing the difference between these line ratio predictions.
Instead, as pointed out by \citet{Zhang2017}, the N2O2 indicator is sensitive to the N/O ratio and temperature variations, which could explain the differences observed between the models.
Also, the discrepancies with our models could arise because we are not considering the impact of post-AGB stars and our dust model is based on the Calzetti law. Moreover, it is important to bear in mind that oxygen abundance indicators involving nitrogen lines may also depend on the N/O ratio evolution in galaxies. In a future development, we plan to include a metal-dependent dust model.

The $\hat{\rm R}$ indicator is a linear combination of R2 and R3 proposed by \citet{Laseter2023}, representing a projection of both indicators to achieve a better fit for low metallicities. As shown in Fig.~\ref {fig:calibrators}, for lower metallicities, our model aligns with the relation presented by \citet{Laseter2023}. However, for the model without dust (blue segmented line), our estimates are shifted downward by approximately $0.08$ dex. This could imply an overestimation of metallicities predicted by our $\hat{\rm R}$ calibration in comparison to those found by \citet{Laseter2023}. 

\begin{figure} 
    \includegraphics[scale=0.6]{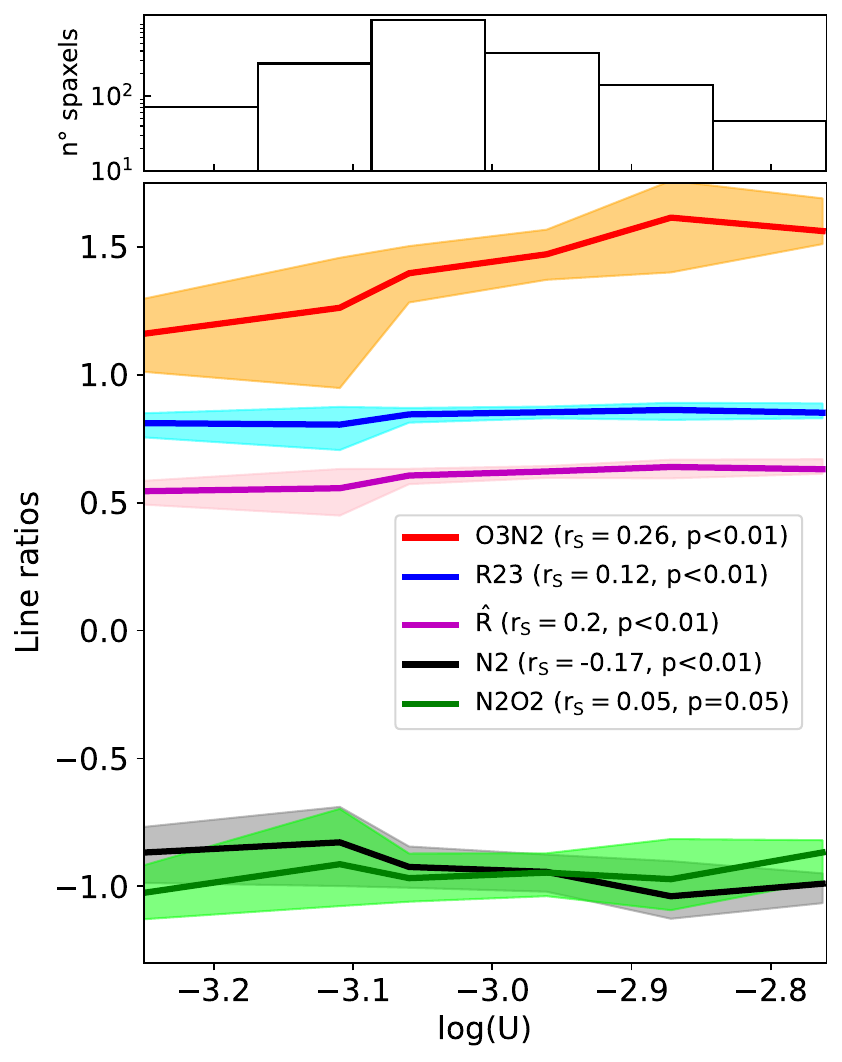}
    \caption{Low panel:Median predicted N2, R23, $\hat{\rm R}$, O3N2 and N2O2 as a function of the ionization parameter. The shaded regions indicate the interquartile range (25$^\text{th}$–75$^\text{th}$ percentiles) of the line ratios in each ionization parameter bin. Additionally, we indicate the Spearman parameter ($\rm r_S$) and p-value for each indicator. Upper panel: distribution of the median log(U)  per spaxel.} 
    \label{fig:ratio_logU} 
\end{figure}

The lower panel of Fig.~\ref{fig:ratio_logU} shows the dependence of  O3N2, R23, $\hat{\rm R}$, N2 and N2O2 on log(U) (bottom panel), considering the non-attenuated spectra by dust along LOS. The tracks represent the median values of the corresponding metallicity indicators per bin of log(U) (each bin contains at least 10 spaxels). The shaded regions represent the interquartile range of the line ratios in each ionization parameter bin, with the lower and upper limits corresponding to the $25^{\rm th}$ and $75^{\rm th}$ percentiles, respectively. The upper panel displays the distribution of the medians of the ionization parameters per spaxel in the simulated sample. Generally, the median simulated log(U) are in good agreement with spatially resolved data, as shown, for instance, by \citet{Mingozzi2020}, \citet{Cameron2021} and \citet{Perez_Montero2023} in the case of star-forming MaNGA galaxies. 

From Fig. \ref{fig:ratio_logU}, we can observe that the N2O2 index (green line) is only weakly sensitive to the ionization parameter. This relation yields a Spearman correlation coefficient of $0.05$ (p $=0.05$). Many authors have reported that this indicator is not strongly dependent on the ionization parameter or variations in the shape of the ionizing spectrum \citep[e.g.,][]{Dopita2000, Dopita2013, Kewley_Dopita2002, Zhang2017}. This makes it a good metallicity indicator even in the presence of DIGs, as we mentioned previously. Additionally, this figure shows an increasing dependence of the O3N2 index on log(U), with a Spearman parameter of $0.26$ (p $<0.01$). In the literature, it has been reported that this dependence on the ionization parameter may contribute to the scatter in the relationship between log(O3N2) with oxygen abundance displayed in Fig. \ref{fig:calibrators} \citep[e.g.,][]{Yin2007, PerezMontero_Contini2009}. This is because a decrease in ionization parameter leads to enhancement of \text{[\ion{N}{ii}]}/H$_\alpha$ and a decrease in \text{[\ion{O}{iii}]}/H$_\beta$ at the same time. This behavior should also imply the existence of an anti-correlation between N2 and log(U) \citep{PerezMontero_Diaz2005}. In our sample we found a Spearman parameter of $-0.17$ with p $<0.01$ between this indicator and log(U), which would confirm a weak anti-correlation between N2 and the ionization parameter. Finally, we found a slight positive dependence of R23 and $\hat{\rm R}$ on ionization parameter, with Spearman parameters of $0.12$ and $0.20$, respectively (p $<0.01$). This indicates that the dependence of $\hat{\rm R}$ on the ionization parameter is stronger than that of R23.

\begin{figure} 
    \hspace*{-0.5cm} 
    \includegraphics[scale=0.5]{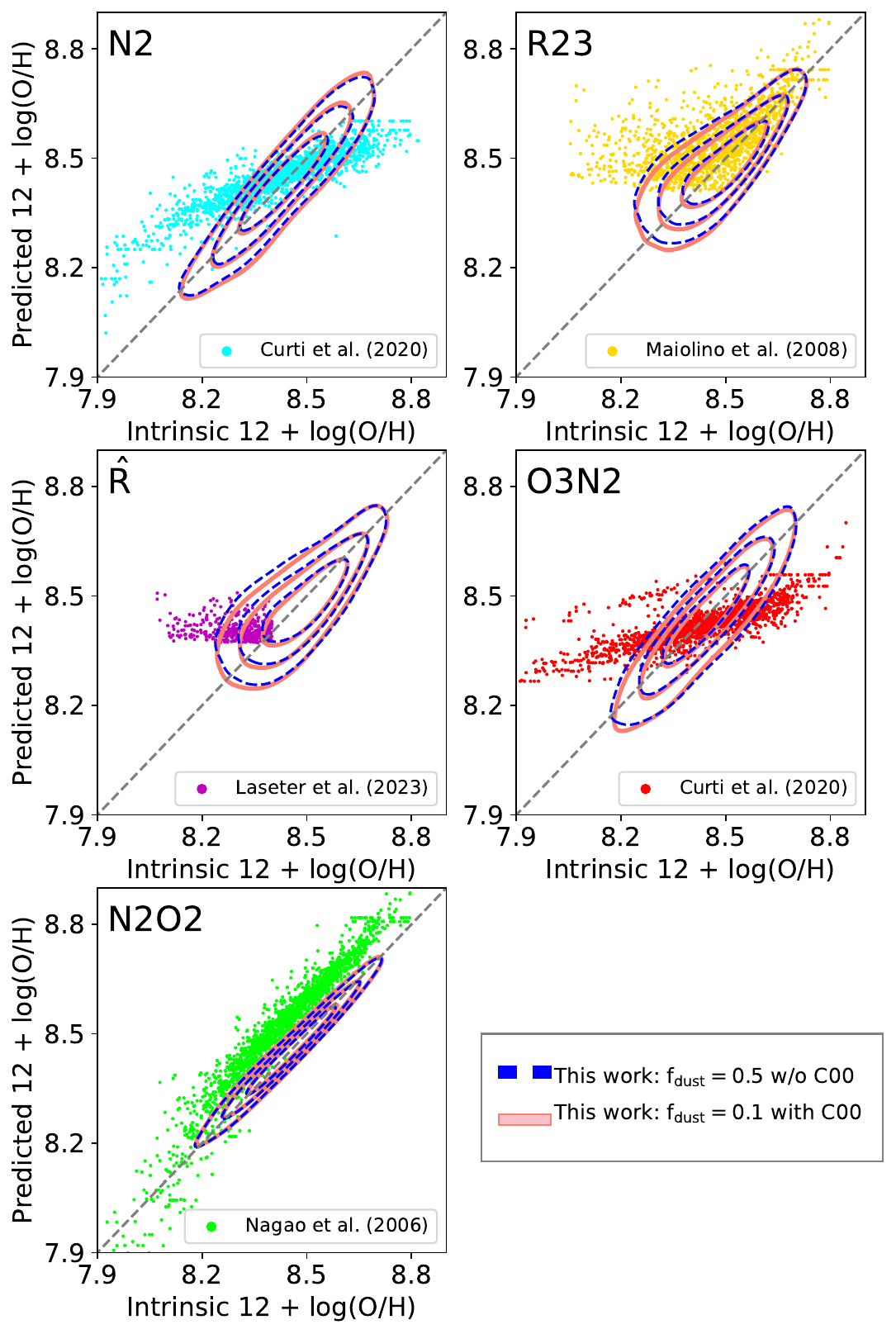} 
    \caption{Comparison between intrinsic gas-phase metallicities and values predicted from PRISMA spectra with f$_\text{dust}=0.5$ (blue contours) and with f$_\text{dust}=0.1$ and the Calzetti law (pink contours). The contours represent the density distributions of spaxels, where the inner, central and outer contours enclose the 30, 50 and 90 percentiles of the spaxels. The predicted abundances from the selected literature calibrations are also displayed (colored dots). The gray, dashed line depicts the 1:1 relation.} 
    \label{fig:true_predicted} 
\end{figure}

Figure~\ref{fig:true_predicted} depicts the predicted versus intrinsic metallicity values for the N2, R23, $\hat{\rm R}$, O3N2 and N2O2 indices, where the blue and pink contours represent the relations obtained by considering the non-attenuated and dust-attenuated spectra along the LOS, respectively. The contours give the density distribution of spaxels, where the inner, central and outer contours enclose 30, 50 and 90 per cent of the spaxels, respectively. For comparison the results obtained with empirical calibrations for N2 and O3N2 \citep[][cyan and red dots]{Curti2020}, R23 \citep[][yellow dots]{Maiolino2008}, $\hat{\rm R}$ \citep[][magenta dots]{Laseter2023} and N2O2 \citep[][green dots]{Nagao2006} are also included. To enhance visualization, the relation $y=x$ is overlaid (gray dashed line). Metallicity calculations were performed by using the applicability ranges of each literature calibration as we did in Fig. \ref{fig:calibrators}. Consequently, in the $\hat{\rm R}$ panel of Fig.~\ref{fig:true_predicted}, fewer metallicities are computed using the \citet{Laseter2023} calibration, as its applicability range is for $\rm 12+log(O/H) < 8.4$.  

From Fig.~\ref{fig:true_predicted} , we also note that the abundances derived using the semi-empirical N2O2 calibration of \citet{Nagao2006} (the lowest panel) are the ones that best align with those predicted by our model. In contrast, the metallicity estimates obtained with the other semi-empirical calibrations do not show the same level of agreement with our predictions. This could be related to the dependence of these indicators on logU, as N2O2 exhibits a lower sensitivity to the ionization parameter, which may contribute to the closer agreement. This implies that, although PRISMA allows the estimation of logU using the intrinsic properties of the simulation, the current model should be considered with caution and is subject to improvements in future work. However, the observed performance confirms the robustness of the N2O2 index in the estimation of metallicities from synthetic spectra of simulations, particularly considering the complexity involved in determining the ionization state of the gas in the star-forming regions that contribute to
such lines ratio.

\begin{figure} 
    \includegraphics[scale=0.47]{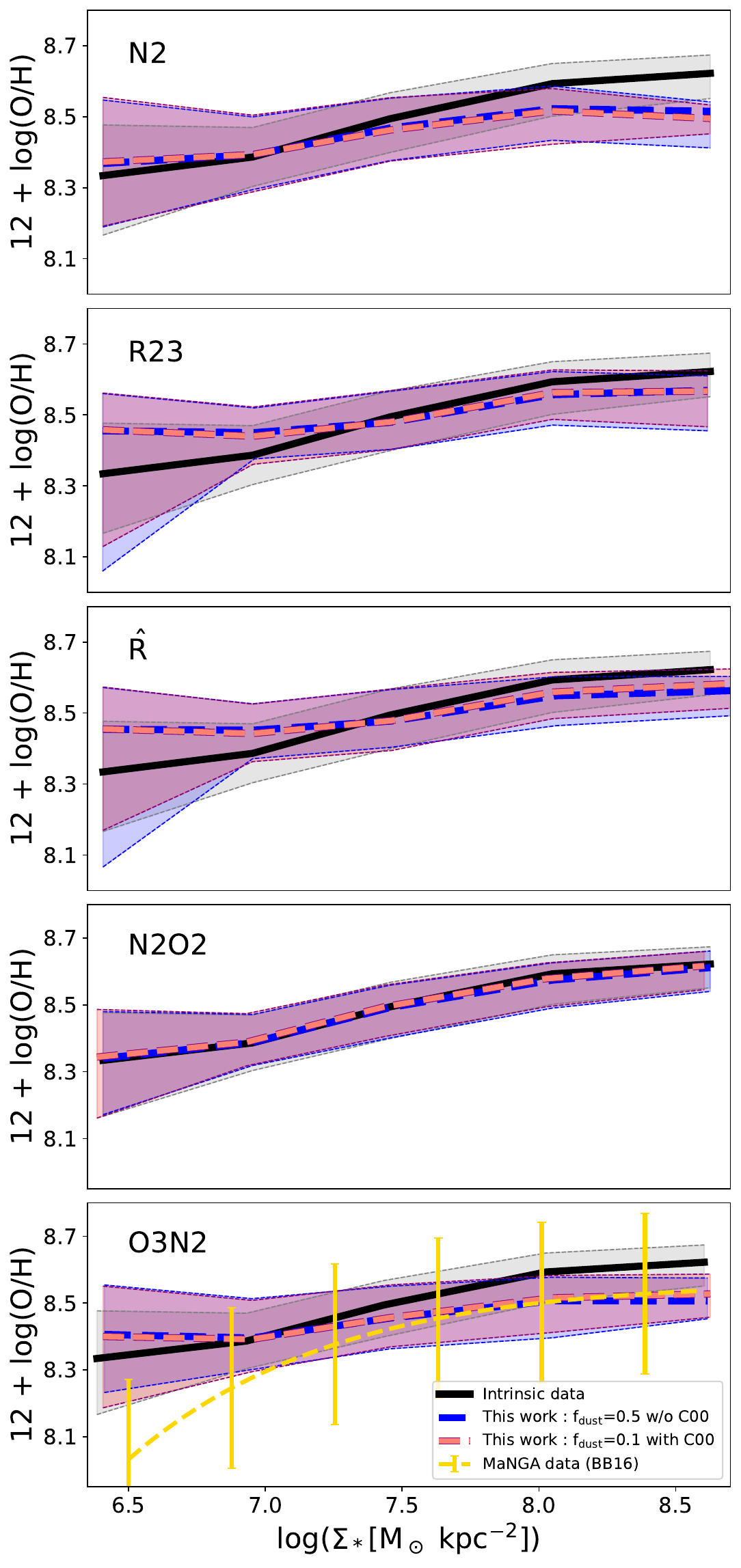} 
    \caption{Resolved Mass-Metallicity relation using intrinsic values from the CIELO simulations (black, solid lines) and the calibrations of N2, R23, $\hat{\rm R}$,  N2O2 and O3N2 provided by PRISMA considering the sample of spectra non-attenuated by ISM dust (blue, dashed lines) and the ISM dust-attenuated spectra considering the Calzetti law (pink, dashed lines). The shaded regions are delimited by the 25-75$^\text{th}$ percentiles. The yellow, dashed line in the O3N2 panel shows the observational results from \citet{Barrera-Ballesteros2016} using MaNGA galaxies (the errobars cores correspond to the standard deviations).} 
    \label{fig:Met_vs_StellarMass} 
\end{figure} 

Finally, the use of our calibrations allowed us to estimate the predicted resolved scaling relation rMZR for our simulated sample. This can be seen in Fig.~\ref{fig:Met_vs_StellarMass}, where  we display the intrinsic (black, solid lines) and the predicted (blue and pink dashed lines for non-attenuated and ISM dust-attenuated spectra, respectively) gas-phase abundances as a function of the surface stellar mass density ($\rm \Sigma_* = M_*/A_\text{spaxel}$) per spaxel. To predict these values, we adopted our calibrations for N2, R23, $\hat{\rm R}$, N2O2 and O3N2. The shaded gray, blue and pink regions depict the respective $25^\text{th}$ and $75^\text{th}$ percentiles of the metallicities\footnote{The median rMZR are estimated in bins of $\rm \Sigma_*$, which  contain at least 10 spaxels.} for the intrinsic and predicted estimations. We display the median relation reported by \citet[][yellow, dashed line]{Barrera-Ballesteros2016}, which utilized disk galaxies from the MaNGA survey and employed the O3N2 index from \citet{Marino2013} to determine abundance estimations.

From Fig. \ref{fig:Met_vs_StellarMass}, we can appreciate that, overall, PRISMA captures the intrinsic rMZR of CIELO for the five indicators studied in this paper. However, we note for  O3N2,  $\hat{\rm R}$, and R23 are the ones that depart the most from the intrinsic rMZR. Nevertheless, these differences are not larger of $\sim 0.15 $dex. Hence, despite using a simple model to simulate \ion{H}{ii} regions and their nebular emission, PRISMA generates spectra that trace the intrinsic properties of the simulations. 

Regarding the case with dust attenuation by using the Calzetti law, in 
Figs.~\ref{fig:calibrators}, \ref{fig:true_predicted} and \ref{fig:Met_vs_StellarMass} we illustrate its impact on the generated spectra (pink curves and contours). From Fig. \ref{fig:calibrators}, we observe that the line ratios most influenced by considering dust effects along the LOS were the O3N2 and N2O2 indicators. However, based on the contours and the blue and pink lines in Figs. \ref{fig:true_predicted} and \ref{fig:Met_vs_StellarMass}, the implementation of dust attenuation using the Calzetti law does not produce notable differences when predicting the metallicities of the simulated sample. In fact, the discrepancy between metallicities obtained from non-attenuated and ISM dust-attenuated spectra using the O3N2 indicator is less than $\sim 0.1$ dex. Additionally, the metallicities predicted by the N2O2 indicator align with the intrinsic metallicities of the simulation, regardless of whether the spectra are non-attenuated or attenuated along the LOS.

However, caution should be exercised when estimating other galaxy properties, where the effects of dust attenuation and re-emission along the LOS may potentially be more significant.

\begin{table}[]
\caption{Predicted metallicity calibrations.}\label{tab: indicators_fit_parameters_clz}
\centering
\begin{tabular}{lllll}
\hline \hline
Line ratio      &     Fit  & Lower curve &  Upper curve
\\ \hline
%N2 clz  
\begin{tabular}[c]{@{}c@{}}N2\\ \end{tabular}& \begin{tabular}[c]{@{}l@{}}$\rm c_0=-21.243$\\ $\rm c_1= 4.007$\\ $\rm c_2 = -0.192$\end{tabular}   & \begin{tabular}[c]{@{}l@{}}$\rm c_0= -21.492$\\ $\rm c_1= 4.028$\\ $\rm c_2 = -0.193$\end{tabular} & \begin{tabular}[c]{@{}l@{}}$\rm c_0= -20.994$\\ $\rm c_1= 3.986$\\ $\rm c_2 =  -0.191$\end{tabular}  \\ \hline
%R23   clz
\begin{tabular}[c]{@{}c@{}}R23\\ \end{tabular}& \begin{tabular}[c]{@{}l@{}}$\rm c_0= -38.566$\\ $\rm c_1= 9.645$\\ $\rm c_2= -0.589$\end{tabular}  & \begin{tabular}[c]{@{}l@{}}$\rm c_0= -38.730$\\ $\rm c_1= 9.659$\\ $\rm c_2= -0.590$\end{tabular} & \begin{tabular}[c]{@{}l@{}}$\rm c_0= -38.403$\\ $\rm c_1= 9.631$\\ $\rm c_2= -0.588$\end{tabular}   \\ \hline
%Rhat clz
\begin{tabular}[c]{@{}c@{}}$\hat{\rm R}$\\ \end{tabular}& \begin{tabular}[c]{@{}l@{}}$\rm c_0= -39.436$\\ $\rm c_1 = 9.810$\\ $\rm c_2= -0.599$\end{tabular} & \begin{tabular}[c]{@{}l@{}}$\rm c_0= -39.574$\\ $\rm c_1= 9.821$\\ $\rm c_2= -0.600$\end{tabular} & \begin{tabular}[c]{@{}l@{}}$\rm c_0= -39.298$\\ $\rm c_1= 9.798$\\ $\rm c_2= -0.599$\end{tabular}   \\ \hline
%O3N2 clz
\begin{tabular}[c]{@{}c@{}}O3N2\\ \end{tabular}& \begin{tabular}[c]{@{}l@{}}$\rm c_0= -20.276$\\ $\rm c_1= 6.312$\\ $\rm c_2= -0.438$\end{tabular}   & \begin{tabular}[c]{@{}l@{}}$\rm c_0= -20.683$\\ $\rm c_1= 6.346$\\ $\rm c_2= -0.440$\end{tabular} & \begin{tabular}[c]{@{}l@{}}$\rm c_0 = -19.870$\\ $\rm c_1 = 6.278$\\ $\rm c_2= -0.436$\end{tabular} \\ \hline
%N2O2 clz
\begin{tabular}[c]{@{}c@{}}N2O2\\ \end{tabular}& \begin{tabular}[c]{@{}l@{}}$\rm c_0= 10.313$\\ $\rm c_1= -3.532$\\ $\rm c_2= 0.264$\end{tabular}    & \begin{tabular}[c]{@{}l@{}}$\rm c_0= 10.199$\\ $\rm c_1= -3.522$\\ $\rm c_2= 0.264$\end{tabular}   & \begin{tabular}[c]{@{}l@{}}$\rm c_0= 10.427$\\ $\rm c_1= -3.541$\\ $\rm c_2= 0.265$\end{tabular}     \\ \hline
\end{tabular}
\tablefoot{As in Table \ref{tab: indicators_fit_parameters}, but considering the spectra attenuated by both dust within the \ion{H}{ii} regions and ISM dust within the LOS assuming the Calzetti law \citep{Calzetti2000}. These curves are represented by the pink, dashed lines in Fig. \ref{fig:calibrators}}.
\end{table}

\section{Conclusions}

We developed a new numerical tool, PRISMA, which mocks IFU-like data cubes of simulated galaxies in order to estimate their physical properties through an observational approach. We used galaxies from the CIELO Project performed with intermediate and high resolution as test-beds. However, PRISMA can be applied to any type of simulations providing the expected parameters. 

For each spaxel, we calculated the synthetic spectra considering the radiation emitted by the stellar populations and the nebular emission by the ionized gas near the youngest stellar populations. A fundamental aspect of PRISMA is that the nebular emission is calculated based on the properties of the gas surrounding the recently born stellar populations  \citep[see][for alternative implementations]{Nanni2022a, Hirschmann2023, Garg2023}. This allowed us to model logU by using the information on the gas component associated with the star forming regions. Although this is a simplified scheme of a \ion{H}{ii} region, it yields a suitable representation of its properties. 

Despite relying on simplified assumptions to model \ion{H}{ii} regions and their nebular emission, our scheme effectively produced spectra that closely recovered the intrinsic SFR and gas-phase metallicities of the galaxies within our sample. Specifically, we employed five metallicity indicators, N2, R23, $\hat{\rm R}$, O3N2, and N2O2, to derive the oxygen abundance, allowing us to compare the effectiveness of each indicator individually. We further analyzed the dependence of these line ratios on both metallicity and the ionization parameter.

Our analysis found an increasing dependence of O3N2, $\hat{\rm R}$, and R23 on logU. Furthermore, we observed no significant correlation between N2 and N2O2 with logU. Additionally, we developed new calibrations for these metallicity indices, which were then applied to derive the rMZR for our sample, yielding results consistent with the intrinsic rMZR.

In addition, we examined the impact of dust attenuation along the line of sight (LOS) on our results. For this, we adopted the Calzetti attenuation law \citep{Calzetti2000} considering E(B-V) = 0.3 mag. From these spectra, we found that the O3N2 and N2O2 indicators are the most affected when applying the Calzetti law. However, when using these indicators to estimate the metallicities of the sample, there were no substantial discrepancies ($< 0.1$ dex) compared to the estimates from spectra without ISM dust attenuation. In a future development, we expect to improve the dust modeling by including a more realistic representation. 

These results confirm the robustness of PRISMA in retrieving the intrinsic properties of simulated galaxies, enabling a better comparison with observational data. In this way, this numerical tool can be used to test the impact of different observational issues such as  dust absorption on the generated spectra.  In a forthcoming paper, we aim to apply this tool to study the resolved metallicity relations.

\begin{acknowledgements}
      We acknowledge the comments of the anonymous referee that contributed to improve this paper. We thanks Stephane Charlot for useful comments and discussion. AC thanks the N\'ucleo Milenio ERIS, Project N°NCN2021\_017.
      ES acknowledges funding by Fondecyt-ANID Postdoctoral 2024 Project N°3240644 and thanks the N\'ucleo Milenio ERIS. 
      PBT acknowledges partial funding by Fondecyt-ANID 1240465/2024, N\'ucleo Milenio ERIS, and ANID Basal Project FB210003. This project has received funding from the European Union Horizon 2020 Research and Innovation Programme under the Marie Sklodowska-Curie grant agreement No 734374- LACEGAL. 
      MB gratefully acknowledges support from the ANID BASAL project FB210003 and from the FONDECYT regular grant 1211000. This work was supported by the French government through the France 2030 investment plan managed by the National Research Agency (ANR), as part of the Initiative of Excellence of Université Côte d’Azur under reference number ANR-15-IDEX-01. 
      PJ acknowledges support form FONDECYT Regular 1231057 and Millennium Nucleus ERIS NCN2021\_017.
      JV acknowledges financial support from the State Agency for Research of the Spanish MCIU through Center of Excellence Severo Ochoa award  to the IAA CEX2021- 001131-S funded by MCIN/AEI/10.13039/501100011033, and from grant PID2022-136598NB-C32 “Estallidos”.
We acknowledge the use of  the Ladgerda Cluster (Fondecyt 1200703/2020 hosted at the Institute for Astrophysics, Chile), the NLHPC (Centro de Modelamiento Matem\'atico, Chile) and the Barcelona Supercomputer Center (Spain). 
\end{acknowledgements}

\bibliographystyle{aa}
\bibliography{bibliography}

\end{document}